\documentstyle[12pt]{article}
\begin{document}

\makeatletter
\@addtoreset{equation}{section}
\def\theequation{\thesection.\arabic{equation}}
\makeatother

\begin{flushright}{December, 1999\\
UT-868}
\end{flushright}
\vskip 0.5 truecm

\begin{center}
{\large{\bf  Note on the Gauge Fixing in Gauge Theory }}
\end{center}
\vskip .5 truecm
\centerline{\bf Kazuo Fujikawa and Hiroaki Terashima}
\vskip .4 truecm
\centerline {\it Department of Physics,University of Tokyo}
\centerline {\it Bunkyo-ku,Tokyo 113,Japan}
\vskip 0.5 truecm

\begin{abstract}
In the absence of Gribov complications, the modified gauge 
fixing in gauge theory\\
$ \int{\cal D}A_{\mu}\{\exp[-S_{YM}(A_{\mu})-\int f(A_{\mu})dx]
/\int{\cal D}g\exp[-\int f(A_{\mu}^{g})dx]\}$\\
for example,  
$f(A_{\mu})=(1/2)(A_{\mu})^{2}$, is identical to the conventional 
Faddeev-Popov formula\\
$\int{\cal D}A_{\mu}\{\delta(D^{\mu}\frac{\delta f(A_{\nu})}{\delta 
A_{\mu}})/\int {\cal D}g\delta(D^{\mu}\frac{\delta f(A_{\nu}^{g})}
{\delta A_{\mu}^{g}})\}\exp[-S_{YM}(A_{\mu})]$\\
if one takes into account the variation of the gauge field along the 
entire gauge orbit. Despite of its quite 
different appearance,the modified formula defines 
a local and BRST invariant theory and thus ensures unitarity at least in
 perturbation theory. In the presence of Gribov
complications, as is expected in non-perturbative Yang-Mills 
theory, the modified formula is equivalent to the 
conventional formula but not identical to it:Both of the 
definitions give rise to non-local theory in general and thus 
the unitarity is not obvious. Implications of the present 
analysis on the lattice regularization are briefly discussed.

\end{abstract}

\section{Introduction}

The standard gauge fixing procedure of general gauge theory
formulated by Faddeev and Popov\cite{faddeev} provides a convenient 
framework
for perturbation theory. The BRST symmetry appearing 
there\cite{brs} 
controls the Slavnov-Taylor identities\cite{slavnov} and ensures the 
renormalizability and unitarity. This formulation however 
suffers from  Gribov complications\cite{gribov} in the non-perturbative
level. The lattice formulation of gauge theory is known to
introduce further complications which may partly be the artifacts 
of lattice regularization. A naive modification of BRST invariant
formulation of continuum theory\cite{fujikawa} does not quite resolve the 
basic issue of lattice regularization, as is illustrated by the no-go 
theorem of Neuberger\cite{neuberger} 
about the lattice implementation of BRST symmetry. Recently the 
issue related to this last point has been partly 
resolved by Testa\cite{testa}. 

A possible generalization of the Faddeev-Popov
formula, which may provide an alternative approach to 
 the Gribov-type complications,
has been proposed by Zwanziger\cite{zwanziger},
and Parrinello and Jona-Lasinio\cite{jona-lasinio}.
It is known that their modified formula reduces to the 
conventional Faddeev-Popov formula in a specific limit of the 
gauge fixing
parameter\cite{zwanziger}\cite{jona-lasinio}\cite{slavnov2}. 
In the present note, we show explicitly that  the modified formula
is reduced to the Faddeev-Popov formula in the absence of 
Gribov complications if one takes into account
the motion of the gauge variable along the entire gauge orbit, 
{\em without} taking a specific limit of the gauge fixing 
parameter.
We thus see that the modified formula, despite of its quite 
different appearance, does not go beyond the Faddeev-Popov
prescription. In particular, the locality and unitarity is 
ensured by both
prescriptions in the absence of Gribov complications, such as in
perturbation theory. In the 
presence of Gribov complications, as is expected in 
non-perturbative formulation, both of the original Faddeev-
Popov formula and the modified formula, which are equivalent 
but not identical to each other, give rise to  non-local 
action: The validity of unitarity is thus not obvious. 

In contrast, a 
modified BRST formula\cite{fujikawa}
ensures unitarity if one assumes the asymptotic condition 
such as the LSZ condition\cite{lsz}, but the full justification 
of 
the modified BRST formulation in the non-perturbative level is 
absent at this moment. 

\section{Non-Abelian gauge theory}
In this note we 
exclusively work on the Euclidean theory, and denote the local 
gauge invariant action by $S_{0}(A_{\mu})$. 
We start with the modified formula\cite{zwanziger}
\cite{jona-lasinio}
\begin{eqnarray}
Z&=&\int{\cal D}A_{\mu}\{e^{-S_{0}(A_{\mu})- \int f(A_{\mu})dx}/\int{\cal 
D}h e^{-\int f(A^{h}_{\mu})dx}\}\nonumber\\
&&=\int{\cal D}A_{\mu}^{\omega}\{e^{-S_{0}(A^{\omega}_{\mu})- \int 
g(A^{\omega}_{\mu})dx}/\int{\cal D}h e^{-\int g(A^{h\omega}_{\mu})dx}\}
\end{eqnarray}
where we suppress the non-Abelian index, and $\omega$ and 
$h$ stand for the gauge parameters; $A^{h\omega}_{\mu}$ stands 
for the field variable obtained
from $A^{\omega}_{\mu}$ by a gauge transformation specified by
$h$.We also defined 
\begin{equation}
g(A^{\omega}_{\mu})\equiv f(A^{\omega}_{\mu})- f(A^{\omega_{0}}_{\mu})
\end{equation}
which satisfies $g(A^{\omega}_{\mu})\geq 0$, where the  
 variable   $A^{\omega_{0}}_{\mu}$ is defined by
\begin{equation}
Min_{\omega}f(A^{\omega}_{\mu}) = f(A^{\omega_{0}}_{\mu}).
\end{equation}
Namely, $\omega_{0}(x)$ is the value of the gauge orbit parameter 
$\omega(x)$ which gives rise to the minimum value of 
$f(A^{\omega}_{\mu})$ at each point of the space-time. We may generally
assume $f(A^{\omega}_{\mu})> 0$ (or more generally, bounded
from below) to
ensure the convergence of the path integral in the denominator
in (2.1). By definition we have
\begin{eqnarray}
\delta_{\omega} \int f(A^{\omega}_{\nu})dx&=& \int dx\frac{\delta 
f(A^{\omega}_{\nu})}{\delta
\omega(x)}\delta\omega(x)\nonumber\\
&=&\int dx\frac{\delta f(A^{\omega}_{\nu})}{\delta
A_{\mu}^{\omega}(x)}D_{\mu}\delta\omega(x)=0 
\end{eqnarray}
for any choice of $\delta\omega(x)$ at $\omega(x)=\omega_{0}(x)$, namely, 
$\omega_{0}(x)$ is implicitly defined by
\begin{equation}
 D_{\mu}\frac{\delta f(A^{\omega_{0}}_{\nu})}{\delta
A_{\mu}^{\omega_{0}}(x)}=0.
\end{equation}
We understand that the absence of Gribov complications implies 
that Eq.(2.5) has a unique solution $\omega_{0}(x)$ for each 
gauge orbit.In this case, Eq.(2.3) also has a unique solution 
$\omega_{0}(x)$: Otherwise, we find multiple solutions for (2.4)
by choosing $\delta\omega(x)$ as a $\delta$-functionally peaked 
function at such a space-time point. 

We next re-write the partition function $Z$ as 
\begin{eqnarray}
Z&=&\int{\cal D}A_{\mu}^{\omega}\{\frac{\int{\cal D}\Lambda\delta(
\tilde{g}(A^{\omega}_{\nu})-\Lambda)e^{-S_{0}(A^{\omega}_{\nu})
-\int\Lambda^{2}dx}}{\int{\cal D}\Lambda^{\prime}{\cal D}h\delta(
\tilde{g}(A^{h\omega}_{\nu})-\Lambda^{\prime})e^{-\int{\Lambda^{\prime}}^
{2}dx}}\}\nonumber\\
&=&\int{\cal D}A_{\mu}^{\omega}\{\frac{\int{\cal 
D}\Lambda\delta(\tilde{g}(A^{\omega}_{\nu}))e^{-S_{0}(A^{\omega}_{\nu})
-\int\Lambda^{2}dx}}{\int{\cal D}\Lambda^{\prime}{\cal D}h\delta(
\tilde{g}(A^{h\omega}_{\nu}))e^{-\int{\Lambda^{\prime}}^{2}dx}}\}
\end{eqnarray}
where we defined 
\begin{eqnarray}
&&\tilde{g}(A^{\omega}_{\nu})=\sqrt{g(A^{\omega}_{\nu})},\ \ \ \  
\tilde{g}(A^{\omega}_{\nu})\geq 0\nonumber\\
&&\tilde{g}(A^{\omega}_{\nu})=-\sqrt{g(A^{\omega}_{\nu})},\ \ \ \ 
\tilde{g}(A^{\omega}_{\nu})< 0
\end{eqnarray}
with $(\tilde{g}(A^{\omega}_{\nu}))^{2}=g(A^{\omega}_{\nu})$.
In the second expression of (2.6) $\omega(x)$ is a {\em generic} gauge 
orbit parameter in the 
infinitesimal neighborhood of $\omega_{0}(x)$, and $h(x)$ is a 
{\em generic}
gauge parameter in the infinitesimal neighborhood of the unit
element. Here we used the fact that we can bring the relation
\begin{equation}
g(A^{\omega}_{\nu})-\Lambda^{2}=0
\end{equation}
to
\begin{equation}
g(A^{\omega^{\prime}}_{\nu})=0
\end{equation}
by choosing a suitable gauge parameter $\omega^{\prime}(x)$ for an 
arbitrary $\Lambda(x)$ in the absence of Gribov 
complications.In fact,$\omega^{\prime}(x)=\omega_{0}(x)$ in the absence
of  Gribov complications. This statement is established by noting
 that we can 
compensate any variation of $\delta\Lambda(x)$ by a suitable change of 
the gauge orbit parameter $\delta\omega(x)$ as 
\begin{equation}
\frac{\delta g(A^{\omega}_{\nu})}{\delta\omega(x)}\delta\omega(x)
=\frac{\delta 
f(A^{\omega}_{\nu})}{\delta\omega(x)}\delta\omega(x)=\delta\Lambda^{2}(x)
\end{equation}
because
\begin{equation}
\frac{\delta f(A^{\omega}_{\nu})}{\delta\omega(x)}\neq0
\end{equation}
for $\omega(x)\neq\omega_{0}(x)$ in the absence of  Gribov 
complications. By a repeated application of infinitesimal gauge 
transformations, we can thus satisfy the 
relation (2.9). We also note that $g(A^{\omega}_{\mu})$ is unbounded from 
above, since otherwise 
the path integral $\int{\cal D}h e^{-\int g(A^{h\omega}_{\mu})dx}$ diverges
in the defining path integral.
We emphasize that we use the {\em generic} parameter
$\omega(x)=\omega_{0}(x)+\delta\omega(x)$ as the gauge parameter for 
the integration variable ${\cal D}A_{\mu}^{\omega}$ in the 
second expression of (2.6);
this fact is important to avoid the appearance of non-trivial
Jacobian which generally arises when we change the path integral 
variable, which satisfies
precisely $\delta(\tilde{g}(A^{\omega}_{\nu})-\Lambda)$, to
$\delta(\tilde{g}(A^{\omega}_{\nu}))$. We are considering the 
functional space of the gauge parameter, and the procedure 
in (2.6) corresponds to a shift of the origin of the functional 
space, which does not give rise to a Jacobian factor. Note that
$\omega_{0}=\omega(A_{\mu}^{phys})$, where 
$A_{\mu}^{phys}$ is
the physical component of $A_{\mu}$ independent of the gauge 
orbit parameter itself.  

We thus finally write the partition function (2.6) after the 
path integration over $\Lambda$ and $\Lambda^{\prime}$ as 
\begin{equation}
Z=\int{\cal D}A_{\mu}^{\omega}\{\delta(
\tilde{g}(A^{\omega}_{\nu}))/\int{\cal D}h\delta(
\tilde{g}(A^{h\omega}_{\nu}))\}
 e^{-S_{0}(A^{\omega}_{\nu})}.
\end{equation}
This is the standard Faddeev-Popov formula for gauge theory with
a specific gauge fixing $\tilde{g}(A^{\omega}_{\nu})=
\sqrt{g(A^{\omega}_{\nu})}=0$.

To write the above path integral formula (2.12) in a more 
manageable manner,
we use a representation of the $\delta$-function
\begin{equation}
\delta(\tilde{g}(A^{\omega}_{\nu}))
\equiv\frac{\lim_{\alpha\rightarrow 0}e^{-
\frac{1}{2\alpha}\int g(A^{\omega}_{\nu})dx}}{\lim_{\alpha\rightarrow 0}
\int{\cal D}\Lambda e^{-\frac{1}{2\alpha}\int dx\Lambda^{2}}}
\end{equation}
and we note the property 
\begin{eqnarray}
g(A^{\omega_{0}+\delta\omega}_{\nu})&=&g(A^{\omega_{0}}_{\nu}) + 
\int dx(\frac{\delta 
f(A^{\omega}_{\nu})}{\delta\omega(x)})_{\omega=\omega_{0}}\delta\omega(x)
\nonumber\\
&&+\int dx dy \delta\omega(x)\{\frac{-\delta}{\delta\omega(y)}[
D_{\mu}(A^{\omega})\frac{\delta f(A^{\omega}_{\nu})}{\delta A_{\mu}^
{\omega}(x)}]\}_{\omega=\omega_{0}}\delta\omega(y)\nonumber\\
&=&\int dx dy \delta\omega(x)\{\frac{-\delta}{\delta\omega(y)}[
D_{\mu}(A^{\omega})\frac{\delta f(A^{\omega}_{\nu})}
{\delta A_{\mu}^
{\omega}(x)}]\}_{\omega=\omega_{0}}\delta\omega(y)
\end{eqnarray}
where we used 
\begin{eqnarray}
&&g(A^{\omega_{0}}_{\nu})=0,\nonumber\\
&&\int dx(\frac{\delta 
f(A^{\omega}_{\nu})}{\delta\omega(x)})_{\omega=\omega_{0}}\delta\omega(x)
=0.
\end{eqnarray}
We emphasize that the limiting expression (2.13)is used as a 
compact 
parametrization of the $\delta$-function, and the relation (2.12)
itself  is established {\em without} taking any limit in the 
gauge 
fixing parameter.

We thus obtain
\begin{eqnarray}
\int{\cal D}\delta\omega
\delta(\tilde{g}(A^{\omega_{0}+\delta\omega}_{\nu}))&=&
\frac{\lim_{\alpha\rightarrow 0}\int{\cal D}\delta\omega e^{-
\frac{1}{2\alpha}\int dx dy 
\delta\omega(x)\{\frac{-\delta}{\delta\omega(y)}[
D_{\mu}(A^{\omega})\frac{\delta f(A^{\omega}_{\nu})}{\delta A_{\mu}^
{\omega}(x)}]\}_{\omega=\omega_{0}}\delta\omega(y)}}
{\lim_{\alpha\rightarrow 0}\int{\cal D}\Lambda e^{-
\frac{1}{2\alpha}\int dx\Lambda^{2}}}\nonumber\\
&=&\frac{1}{det \sqrt{O}}
\end{eqnarray}
where we defined
\begin{equation}
O(x,y)\equiv \{ \frac{-\delta}{\delta\omega(y)}[
D_{\mu}(A^{\omega})\frac{\delta f(A^{\omega}_{\nu})}{\delta A_{\mu}^
{\omega}(x)}] \}_{\omega=\omega_{0}}
\end{equation}
and we also  defined an operator $\sqrt{O}$ which 
satisfies
\begin{equation}
det \sqrt{O}= \sqrt{det O}.
\end{equation}
Note that we may assume that the operator $O(x,y)$ is proportional
to $\delta(x-y)$ and a positive definite operator  in the 
absence of Gribov complications: We can thus define 
$\sqrt{O}$ in (2.18) as an operator whose eigenvalues are given 
by the square root of the eigenvalues of $O$ and still 
proportional to $\delta(x-y)$. 

We also have
\begin{eqnarray}
&&\delta(\tilde{g}(A^{\omega_{0}+\delta\omega}_{\nu}))
=\frac{\lim_{\alpha\rightarrow 0} e^{-
\frac{1}{2\alpha}\int dx dy 
\delta\omega(x)O(x,y)\delta\omega(y)}}{\lim_{\alpha\rightarrow 
0}\int{\cal D}\Lambda e^{-
\frac{1}{2\alpha}\int dx\Lambda^{2}}}\nonumber\\
&=&\frac{\lim_{\alpha\rightarrow 0} e^{-
\frac{1}{2\alpha}\int dx dy D_{\mu}(\frac{\delta f(A^{\omega_{0}+
\delta\omega}_{\nu})}{\delta 
A_{\mu}^{\omega_{0}+\delta\omega}})(y)O(x,y)^{-1}D_{\lambda}(\frac{\delta 
f(A^{\omega_{0}+
\delta\omega}_{\nu})}{\delta 
A_{\lambda}^{\omega_{0}+\delta\omega}})(x)}}{\lim_{\alpha\rightarrow 
0}\int{\cal D}\Lambda e^{-
\frac{1}{2\alpha}\int dx\Lambda^{2}}}\nonumber\\
&=& \delta (\frac{1}{\sqrt{O}}D_{\mu}(\frac{\delta 
f(A^{\omega_{0}+\delta\omega}_{\nu})}{\delta 
A_{\mu}^{\omega_{0}+\delta\omega}}))
\end{eqnarray}
where we used the fact that  
\begin{eqnarray}
D_{\mu}(\frac{\delta f(A^{\omega_{0}+\delta\omega}_{\nu})}{\delta 
A_{\mu}^{\omega_{0}+\delta\omega}})(x) &=& D_{\mu}(\frac{\delta 
f(A^{\omega_{0}}_{\nu})}{\delta A_{\mu}^{\omega_{0}}})(x)- 
\int dyO(x,y)\delta\omega(y)\nonumber\\
&=&-\int dyO(x,y)\delta\omega(y).
\end{eqnarray}
In the absence of  Gribov complications, we thus have
\begin{eqnarray}
Z&=&\int{\cal D}A_{\mu}^{\omega}\delta 
(\frac{1}{\sqrt{O}}D_{\mu}(\frac{\delta f(A^{\omega}_{\nu})}{\delta 
A_{\mu}^{\omega}}))det\sqrt{O}e^{-S_{0}(A^{\omega}_{\mu})}\nonumber\\
&&=\int{\cal D}A_{\mu}^{\omega}{\cal D}B{\cal D}\bar{c}{\cal D}c
e^{\int[-iB\frac{1}{\sqrt{O}}D_{\mu}(\frac{\delta f(A^{\omega}_{\nu})}{\delta 
A_{\mu}^{\omega}})+\bar{c}\sqrt{O}c]dx-S_{0}(A^{\omega}_{\mu})}
\nonumber\\
&&=\int{\cal D}A_{\mu}^{\omega}{\cal D}B{\cal D}\bar{c}{\cal D}c
e^{\int[-iB D_{\mu}(\frac{\delta f(A^{\omega}_{\nu})}{\delta A_{\mu}^{\omega}})
+\bar{c}Oc]dx-S_{0}(A^{\omega}_{\mu})}    
\end{eqnarray}
where $c$ and $\bar{c}$ stand for the Faddeev-Popov ghosts, and 
$B$ is the Nakanishi-Lautrup field. In the last expression we 
re-defined the {\em auxiliary} 
variables as 
\begin{eqnarray}
B\rightarrow B\sqrt{O}\nonumber\\
\bar{c}\rightarrow \bar{c}\sqrt{O}
\end{eqnarray}
which leaves the path integral measure invariant. 
  
We have thus established the equality
\begin{eqnarray}
&&\int{\cal D}A_{\mu}\{e^{-S_{0}(A_{\mu})- \int f(A_{\mu})dx}/\int{\cal 
D}h e^{-\int f(A^{h}_{\mu})dx}\}\nonumber\\
&&=\int{\cal D}A_{\mu}{\cal D}B{\cal D}\bar{c}{\cal D}c
e^{\int[-iB D_{\mu}(\frac{\delta f(A_{\nu})}{\delta 
A_{\mu}})+\bar{c}Oc]dx-S_{0}(A_{\mu})}    
\end{eqnarray}     
up to a field re-definition of auxiliary variables in 
(2.22).\footnote{What we have 
done in the process
from (2.12) to (2.23) is illustrated in the case of an ordinary 
function as follows: Consider $f(x)\simeq
f(x_{0})+f^{\prime}(x_{0})(x-x_{0})+\frac{1}{2!}f^{\prime\prime}(x_{0})(x
-x_{0})^{2}=
\frac{1}{2!}f^{\prime\prime}(x_{0})(x-x_{0})^{2}$ with $f(x_{0})=
f^{\prime}(x_{0})=0$ and $f^{\prime\prime}(x_{0})>0$. We then 
have $\tilde{f}(x)=\sqrt{f^{\prime\prime}(x_{0})/2}(x-x_{0})$ and
 $\delta(\tilde{f}(x)) \sim 
\delta(\sqrt{f^{\prime\prime}(x_{0})}(x-x_{0}))\sim
\delta(\frac{1}{\sqrt{f^{\prime\prime}(x_{0})}}f^{\prime}(x))$. 
Consequently, $\delta(\tilde{f}(x))/\int dx\delta(\tilde{f}(x))=
\delta(\frac{1}{\sqrt{f^{\prime\prime}(x_{0})}}f^{\prime}(x))
/\int
dx\delta(\sqrt{f^{\prime\prime}(x_{0})}(x-x_{0}))
=\delta(\frac{1}{\sqrt{f^{\prime\prime}(x_{0})}}
f^{\prime}(x))/(\frac{1}{\sqrt{f^{\prime\prime}(x_{0})}})=
\delta(f^{\prime}(x))f^{\prime\prime}(x_{0})$.}

The final path integral formula (2.23) is local and invariant 
under the 
BRST transformation with a Grassmann parameter $\lambda$
\begin{eqnarray}
&&\delta A^{a}_{\mu}=i\lambda(D_{\mu}c)^{a}\nonumber\\
&&\delta c^{a}=-\frac{i\lambda}{2}f^{abc}c^{b}c^{c}\nonumber\\
&&\delta\bar{c}^{a}=\lambda B^{a}\nonumber\\
&&\delta B^{a}=0
\end{eqnarray}
for a rather general
class of gauge fixing function $f(A_{\mu})$.We here write the 
gauge index explicitly. This BRST symmetry
itself holds even before the field re-definition in (2.21), since
the field re-definition (2.22) is consistent with BRST symmetry. 
Eq.(2.23) is the main result of the present note.

So far we assumed the absence of Gribov complications, and thus the
arguments are applicable to perturbation theory. 
In the presence of Gribov complications, as is expected in 
non-perturbative formulation, the analysis becomes 
more complicated. We here understand the Gribov complications 
as simply meaning the appearance of multiple solutions of
\begin{equation}
\delta_{\omega}\int f(A^{\omega}_{\nu})dx=0
\end{equation}
at $\omega(x)=\omega_{k}(x), \ \ k=1,2,...,n$.We also assume 
that these
$\omega_{k}(x)$ are globally defined in the entire space-time.

In the presence of Gribov complications, we cannot make a definite
statement. In the following, we briefly sketch how one can 
transform (2.1) to an expression which is as local as possible; 
this may be relevant for the analysis of the issues related to 
unitarity. In the modified path integral formula\cite{zwanziger}
\cite{jona-lasinio}, the local minimum solutions of (2.25)
correspond to the so-called Gribov copies. The local maximum 
solutions of (2.25) correspond to the so-called Gribov horizons, 
namely, the obstruction in the analysis of (2.10). We can then
still arrive at the second expression in (2.6), but now 
$\omega_{0}(x)$ stands for one of those $\omega_{k}(x)$ which 
give the local minima of $f(A^{\omega})$.  
For these solutions , the operator in (2.17) is considered as 
a positive operator. We thus obtain
\begin{eqnarray}
&&\int{\cal D}A^{\omega}_{\mu}\{e^{-S_{0}(A^{\omega}_{\mu}) 
-\int f(A^{\omega}_{\mu})dx}
/\int{\cal D}h e^{-\int f(A^{h\omega}_{\mu})dx}\}\nonumber\\
&&=\int{\cal D}A^{\omega}_{\mu}\{e^{-S_{0}(A^{\omega}_{\mu}) 
-\int g(A^{\omega}_{\mu})dx}/\int{\cal D}h e^{-\int 
g(A^{h\omega}_{\mu})dx}\}\nonumber\\  
&&=\sum_{k}\int{\cal D}A^{\omega_{k}}_{\nu}\{
\delta(\tilde{g}(A^{\omega_{k}}_{\nu}))/
[\sum_{k}\int{\cal D}h\delta(\tilde{g}(A^{h\omega_{k}}_{\nu}))]\}
e^{-S_{0}(A^{\omega_{k}}_{\nu})}\nonumber\\
&&=\sum_{k}\int{\cal D}A^{\omega_{k}}_{\mu}
\{\delta (\frac{1}{\sqrt{O_{k}}}D_{\mu}
(\frac{\delta f(A^{\omega_{k}}_{\nu})}{\delta A_{\mu}^{\omega_{k}}})) 
/[\sum_{k}\frac{1}{det \sqrt{O_{k}}}]\}e^{-S_{0}
(A^{\omega_{k}}_{\mu})} \nonumber\\
&&=\sum_{k}\int{\cal D}A^{\omega_{k}}_{\mu}\{
\delta (D_{\mu}
(\frac{\delta f(A^{\omega_{k}}_{\nu})}{\delta A_{\mu}^{\omega_{k}}}))det 
\sqrt{O_{k}}
\} /\{\sum_{k}\frac{1}{det 
\sqrt{O_{k}}}\}e^{-S_{0}(A^{\omega_{k}}_{\mu})}
\end{eqnarray}
where the integration variable $A^{\omega_{k}}_{\mu}$ and the variable 
inside the action and the $\delta$-functional constraint stand
for a generic gauge field in the infinitesimal neighborhood of
the local minimum solutions of (2.25).\footnote{We here 
take a view that we should sum over all the Gribov copies, as is 
the case in the modified BRST formulation\protect\cite{lee}. 
It is 
argued in\protect\cite{zwanziger}\protect\cite{jona-lasinio} 
that the integrated 
functional 
 has an absolute minimum solution for 
$Min_{\omega}\int f(A^{\omega}_{\mu})dx$
if one chooses
a suitable $f(A^{\omega}_{\mu})$ such as $f(A^{\omega}_{\mu})=
(1/2)(A^{\omega}_{\mu})^{2}$. This suggests that, if one could 
argue that the Gribov horizon is overcome in the analysis of 
(2.10) in the modified Faddeev-Popov formula, one would have a 
chance  to achieve a relaxation to the absolute minimum of 
$\int f(A^{\omega}_{\mu})dx$ in the path integral. In such a 
case (and if the absolute minimum is unique),the Gribov 
complications would largely be resolved in the 
modified Faddeev-Popov formula. }

On the other hand, the conventional Faddeev-Popov formula gives (
by remembering that the $\delta$-function is by definition 
positive)
\begin{eqnarray}
Z&=&\int{\cal D}A_{\mu}^{\omega}\{\delta (D_{\mu}(\frac{\delta 
f(A^{\omega}_{\nu})}{\delta A_{\mu}^{\omega}}))/\int{\cal D}h \delta 
(D_{\mu}(\frac{\delta f(A^{h\omega}_{\nu})}{\delta A_{\mu}^{h\omega}}))\}
e^{-S_{0}(A^{\omega}_{\mu})}\nonumber\\
&&=\int{\cal D}A_{\mu}^{\omega}\{\delta (D_{\mu}(\frac{\delta 
f(A^{\omega}_{\nu})}{\delta A_{\mu}^{\omega}}))\}/\{\sum_{k}\frac{1}{|det O_{k}
|}\}e^{-S_{0}(A^{\omega}_{\mu})}
\end{eqnarray}
where the summation is over all the solutions of (2.25).

Both of the formulas (2.26) and (2.27) give rise to a 
{\em positive definite} integrand in 
Euclidean theory, but not quite identical to each other. 
(We note that (2.26) and (2.27) are identical in the absence of
Gribov copies.)
However, they are equivalent to each other  in the sense that
both are defined by extracting the gauge volume element from the 
naive path integral measure ${\cal D}A_{\mu}$. Both formulas 
give rise
to non-local theory, and thus the unitarity is not obvious.

In contrast, the BRST invariant ansatz suggested in 
Ref.\cite{fujikawa}
corresponds to a local BRST invariant expression
\begin{eqnarray}
&&\int{\cal D}A_{\mu}^{\omega}\delta (D_{\mu}(\frac{\delta f(A^{\omega}_{\nu})}
{\delta A_{\mu}^{\omega}}))det\{\frac{\delta}{\delta\omega}
[D_{\mu}(\frac{\delta f(A^{\omega}_{\nu})}{\delta A^{\omega}_{\mu}})]\}
e^{-S_{0}(A^{\omega}_{\mu})}\nonumber\\
&&=\int{\cal D}A_{\mu}^{\omega}{\cal D}B{\cal D}\bar{c}{\cal D}c
e^{\int[-iB D_{\mu}(\frac{\delta f(A^{\omega}_{\nu})}{\delta A_{\mu}^{\omega}})
+\bar{c}\frac{\delta}{\delta\omega}
D_{\mu}(\frac{\delta f(A^{\omega}_{\nu})}{\delta A^{\omega}_{\mu}}) 
c]dx-S_{0}(A^{\omega}_{\mu})}    
\end{eqnarray}
where the integrand is {\em no more} positive definite in the presence
of Gribov copies, as we do not take the absolute value of 
$det\{\frac{\delta}{\delta\omega}
[D_{\mu}(\frac{\delta f(A^{\omega}_{\nu})}{\delta A^{\omega}_{\mu}})]\}
$. In (2.28) we integrate over all the gauge field configurations.
The BRST symmetry
combined with the asymptotic condition such as the LSZ prescription
defines a unitary theory. We emphasize that the Lagrangian for the
Faddeev-Popov ghosts in (2.28) is {\em not} degenerate 
in general with respect to the 
degenerate solutions of $D_{\mu}(\frac{\delta f(A^{\omega}_{\nu})}{\delta
A_{\mu}^{\omega}})=0$;consequently, the asymptotic condition such 
as the LSZ condition may well pick up a {\em unique} asymptotic 
field
$A_{\mu}$ despite the presence of the Gribov copies.(In pure 
Yang-Mills theory without the Higgs mechanism, we expect the 
gluon confinement and thus the asymptotic condition may be
replaced by the use of a Wilson loop, for example).

\section{Abelian example}
An example of  Abelian gauge theory
may be illustrative, since we can then work out everything 
explicitly. Note that there is no Gribov complications
in the Abelian theory at least in a continuum formulation. 
 As a simple and useful example, we choose
the gauge fixing function\cite{zwanziger}\cite{jona-lasinio}
\begin{equation}
f(A)\equiv \frac{1}{2}A_{\mu}A_{\mu}
\end{equation}
and 
\begin{equation}
D_{\mu}(\frac{\delta f}{\delta A_{\mu}})=\partial_{\mu}A_{\mu}.
\end{equation}
Our analysis in (2.23) suggests the relation
\begin{eqnarray}
Z&=&\int {\cal D}A^{\omega}_{\mu}\{e^{-S_{0}(A^{\omega}_{\mu})-
\int dx \frac{1}{2}(A^{\omega}_{\mu})^{2}}/\int {\cal D}h e^{-\int dx 
\frac{1}{2}(A^{h\omega}_{\mu})^{2}}\}\nonumber\\
&=&\int {\cal D}A^{\omega}_{\mu}{\cal D}B{\cal D}\bar{c}{\cal D}c 
e^{-S_{0}(A^{\omega}_{\mu})+ \int[-iB\partial_{\mu}A^{\omega}_{\mu}+ 
\bar{c}(-\partial_{\mu}\partial_{\mu})c]dx }.
\end{eqnarray}
To establish this result, we first evaluate 
\begin{eqnarray}
&&\int {\cal D}h e^{-\int dx \frac{1}{2}(A^{h\omega}_{\mu})^{2}} 
\nonumber\\
&&=\int {\cal D}h e^{-\int dx \frac{1}{2}(A^{\omega}_{\mu}+\partial_{\mu}
h)^{2}} \nonumber\\
&&=\int {\cal D}h e^{-\int dx 
\frac{1}{2}[(A^{\omega}_{\mu})^{2}-2(\partial_{\mu}A^{\omega}_{\mu})h + 
h(-\partial_{\mu}\partial_{\mu})h]} 
\nonumber\\
&&=\int {\cal D}B\frac{1}{det\sqrt{-\partial_{\mu}\partial_{\mu}}} 
e^{-\int dx \frac{1}{2}[(A^{\omega}_{\mu})^{2}-2(\partial_{\mu}A^{\omega}
_{\mu})\frac{1}{\sqrt{-\partial_{\mu}\partial_{\mu}}}B + 
B^{2}]}\nonumber\\
&&=\frac{1}{det \sqrt{-\partial_{\mu}\partial_{\mu}}}e^{-\int dx \frac{1}
{2}(A^{\omega}_{\mu})^{2}+\frac{1}{2}\int\partial_{\mu}
A^{\omega}_{\mu}\frac{1}{-\partial_{\mu}\partial_{\mu}}
\partial_{\nu}A^{\omega}_{\nu}dx} 
\end{eqnarray}
where we defined $\sqrt{-\partial_{\mu}\partial_{\mu}}h=B$. Thus
\begin{eqnarray}
Z&=&\int{\cal D}A^{\omega}_{\mu}\{det 
\sqrt{-\partial_{\mu}\partial_{\mu}}\}e^{-S_{0}(A^{\omega}_{\mu})-\frac{1
}{2}\int\partial_{\mu}A^{\omega}_{\mu}\frac{1}{-\partial_{\mu}
\partial_{\mu}}
\partial_{\nu}A^{\omega}_{\nu}dx}
\nonumber\\
&=&\int{\cal D}A^{\omega}_{\mu}{\cal D}B{\cal D}\bar{c}{\cal D}c
e^{-S_{0}(A^{\omega}_{\mu})-\frac{1}{2}\int B^{2}dx +\int
[-iB\frac{1}{\sqrt{-\partial_{\mu}\partial_{\mu}}}\partial_{\mu}A^{\omega
}_{\mu}+\bar{c}\sqrt{-\partial_{\mu}\partial_{\mu}}c]dx}
\end{eqnarray}
which is invariant under the BRST transformation
\begin{eqnarray}
&&\delta A_{\mu}^{\omega}=i\lambda\partial_{\mu}c\nonumber\\
&&\delta c=0\nonumber\\
&&\delta\bar{c}=\lambda B\nonumber\\
&&\delta B=0
\end{eqnarray}
with a Grassmann parameter $\lambda$.Note the appearance of the
imaginary factor $i$ in the term
$iB\frac{1}{\sqrt{-\partial_{\mu}\partial_{\mu}}}\partial_{\mu}A^{\omega}
_{\mu}$ 
in (3.5).

When one defines
\begin{equation}
Z(\alpha)=\int{\cal D}A^{\omega}_{\mu}{\cal D}B{\cal D}\bar{c}{\cal D}c 
e^{-S_{0}(A^{\omega})-\frac{\alpha}{2}\int B^{2}dx +
\int[-iB\frac{1}{\sqrt{-\partial_{\mu}\partial_{\mu}}}\partial_{\mu}
A^{\omega}_{\mu} +\bar{c}\sqrt{-\partial_{\mu}\partial_{\mu}}c]dx}
\end{equation}
one can show that 
\begin{eqnarray}
&&Z(\alpha-\delta\alpha)\nonumber\\
&&=\int{\cal D}A^{\omega}_{\mu}{\cal D}B{\cal D}\bar{c}{\cal D}c 
e^{-S_{0}(A^{\omega}_{\mu})-\frac{\alpha-\delta\alpha}{2}\int B^{2}dx 
+\int[-iB\frac{1}{\sqrt{-\partial_{\mu}\partial_{\mu}}}\partial_{\mu}
A^{\omega}_{\mu}+\bar{c}\sqrt{-\partial_{\mu}\partial_{\mu}}c]dx}\nonumber\\
&&=Z(\alpha)\nonumber\\
&+&\int{\cal D}A^{\omega}_{\mu}{\cal D}B{\cal D}\bar{c}{\cal D}c 
(\frac{\delta\alpha}{2}\int 
B^{2}dx)e^{-S_{0}(A^{\omega}_{\mu})-\frac{\alpha}{2}\int B^{2}dx 
+\int[-iB\frac{1}{\sqrt{-\partial_{\mu}\partial_{\mu}}}\partial_{\mu}
A^{\omega}_{\mu}+\bar{c}\sqrt{-\partial_{\mu}\partial_{\mu}}c]dx}.
\end{eqnarray}
On the other hand,the BRST invariance of the path integral 
measure and the effective action in the exponential factor gives
rise to  
\begin{eqnarray}
&&\int{\cal D}A^{\omega}_{\mu}{\cal D}B{\cal D}\bar{c}{\cal D}c 
\int(\bar{c}B)dxe^{-S_{0}(A^{\omega}_{\mu})-\frac{\alpha}{2}\int B^{2}dx 
+\int[-iB\frac{1}{\sqrt{-\partial_{\mu}\partial_{\mu}}}\partial_{\mu}
A^{\omega}_{\mu}+\bar{c}\sqrt{-\partial_{\mu}\partial_{\mu}}c]dx}\nonumber\\
&&=\int{\cal D}A^{\prime\omega}_{\mu}{\cal D}B^{\prime}{\cal 
D}\bar{c}^{\prime}{\cal D}c^{\prime} 
\int(\bar{c}^{\prime}B^{\prime})dx\nonumber\\
&&\times e^{-S_{0}(A^{\prime\omega}_{\mu})-\frac{\alpha}{2}\int 
{B^{\prime}}^{2}dx 
+\int[-iB^{\prime}\frac{1}{\sqrt{-\partial_{\mu}\partial_{\mu}}}\partial_
{\mu}A^{\prime\omega}_{\mu}+\bar{c}^{\prime}\sqrt{-\partial_{\mu}\partial
_{\mu}}c^{\prime}]dx}\nonumber\\
&&=\int{\cal D}A^{\omega}_{\mu}{\cal D}B{\cal D}\bar{c}{\cal D}c 
\int(\bar{c}B + \delta_{BRST}(\bar{c}B))dx\nonumber\\
&&\times e^{-S_{0}(A^{\omega}_{\mu})-\frac{\alpha}{2}\int B^{2}dx 
+\int[-iB\frac{1}{\sqrt{-\partial_{\mu}\partial_{\mu}}}\partial_{\mu}
A^{\omega}_{\mu}+\bar{c}\sqrt{-\partial_{\mu}\partial_{\mu}}c]dx}
\end{eqnarray}
where the BRST transformed variables are defined by 
$A^{\prime\omega}_{\mu}=A^{\omega}_{\mu}+
i\lambda\partial_{\mu}c, B^{\prime}=B,\bar{c}^{\prime}=\bar{c}+\lambda B, 
c^{\prime}=c $.
Namely, the BRST exact quantity vanishes as 
\begin{eqnarray}
&&\int{\cal D}A^{\omega}_{\mu}{\cal D}B{\cal D}\bar{c}{\cal D}c 
(\int\delta_{BRST}(\bar{c}B)dx)e^{-S_{0}(A^{\omega}_{\mu})-\frac{\alpha}{
2}\int B^{2}dx 
+\int[-iB\frac{1}{\sqrt{-\partial_{\mu}\partial_{\mu}}}\partial_{\mu}
A^{\omega}_{\mu}+\bar{c}\sqrt{-\partial_{\mu}\partial_{\mu}}c]dx}\nonumber\\
&&=
\int{\cal D}A^{\omega}_{\mu}{\cal D}B{\cal D}\bar{c}{\cal D}c 
(\lambda\int B^{2}dx)e^{-S_{0}(A^{\omega})-\frac{\alpha}{2}\int B^{2}dx 
+\int[-iB\frac{1}{\sqrt{-\partial_{\mu}\partial_{\mu}}}\partial_{\mu}
A^{\omega}_{\mu}+\bar{c}\sqrt{-\partial_{\mu}\partial_{\mu}}c]dx}\nonumber\\
&&=0.
\end{eqnarray}
Thus from (3.8) we have the relation
\begin{equation}
Z(\alpha-\delta\alpha)=Z(\alpha).
\end{equation}
Namely, $Z(\alpha)$ is independent of $\alpha$,
and $Z(1)=Z(0)$
. We thus obtain 
\begin{eqnarray}
Z&=&\int{\cal D}A^{\omega}_{\mu}\{e^{-S_{0}(A^{\omega}_{\mu})-\frac{1}{2}
\int(A^{\omega}_{\mu})^{2}dx}/\int{\cal D}h e^{-\frac{1}{2}
\int(A^{h\omega}_{\mu})^{2}dx}\}\nonumber\\
&=&\int{\cal D}A^{\omega}_{\mu}{\cal D}B{\cal D}\bar{c}{\cal D}c 
e^{-S_{0}(A^{\omega}_{\mu})+\int[-iB\frac{1}{\sqrt{-\partial_{\mu}
\partial_{\mu}}}\partial_{\mu}A^{\omega}_{\mu}+\bar{c}\sqrt{-\partial_{\mu}
\partial_{\mu}}c]dx}\nonumber\\
&=&\int{\cal D}A^{\omega}_{\mu}{\cal D}B{\cal D}\bar{c}{\cal D}c 
e^{-S_{0}(A^{\omega}_{\mu})+\int[-iB\partial_{\mu}A^{\omega}_{\mu}+\bar{c
}(-\partial_{\mu}\partial_{\mu})c]dx}
\end{eqnarray}
after a re-definition  of the {\em auxiliary} variables
\begin{eqnarray}
&&B\rightarrow B\sqrt{-\partial_{\mu}\partial_{\mu}}\nonumber\\
&&\bar{c}\rightarrow \bar{c}\sqrt{-\partial_{\mu}\partial_{\mu}} 
\end{eqnarray}
which is consistent with BRST symmetry and leaves the path 
integral measure invariant. We thus established the desired 
result (3.3). 

An alternative way to arrive at the result (3.12) from (3.5) is 
to rewrite the expression (3.5) as
\begin{eqnarray}
&&\int{\cal D}A^{\omega}_{\mu}{\cal D}B{\cal D}\Lambda{\cal 
D}\bar{c}{\cal D}c 
\delta(\frac{1}{\sqrt{-\partial_{\mu}\partial_{\mu}}}
\partial_{\mu}A^{\omega}_{\mu}-\Lambda)e^{-S_{0}(A^{\omega}_{\mu})
-\frac{1}{2}\int(B^{2}+2i\Lambda B)dx 
+\int\bar{c}\sqrt{-\partial_{\mu}\partial_{\mu}}cdx}\nonumber\\
&&=\int{\cal D}A^{\omega}_{\mu}{\cal D}\Lambda{\cal D}\bar{c}{\cal D}c 
\delta(\frac{1}{\sqrt{-\partial_{\mu}\partial_{\mu}}}\partial_{\mu}
A^{\omega}_{\mu}-\Lambda)e^{-S_{0}(A^{\omega}_{\mu})-\frac{1}{2}
\int\Lambda^{2}dx +\int\bar{c}\sqrt{-\partial_{\mu}\partial_{\mu}}cdx}.
\end{eqnarray}
We next note that we can compensate any variation of 
$\delta\Lambda$ by a suitable change of gauge parameter 
$\delta\omega$ inside the $\delta$-function as 
\begin{equation}
\frac{1}{\sqrt{-\partial_{\mu}\partial_{\mu}}}\partial_{\mu}\partial_{\mu
}\delta\omega=\delta\Lambda.
\end{equation}
By a repeated application of infinitesimal gauge transformations 
combined with the invariance of the path integral measure under 
these gauge transformations, we can re-write the formula (3.14) as 
\begin{eqnarray}
&&\int{\cal D}A^{\omega}_{\mu}{\cal D}\Lambda{\cal D}\bar{c}{\cal D}c 
\delta(\frac{1}{\sqrt{-\partial_{\mu}\partial_{\mu}}}\partial_{\mu}
A^{\omega}_{\mu})e^{-S_{0}(A^{\omega}_{\mu})-\frac{1}{2}\int\Lambda^{2}dx 
+\int\bar{c}\sqrt{-\partial_{\mu}\partial_{\mu}}cdx}\nonumber\\
&&=\int{\cal D}A^{\omega}_{\mu}{\cal D}\bar{c}{\cal D}c 
\delta(\frac{1}{\sqrt{-\partial_{\mu}\partial_{\mu}}}\partial_{\mu}
A^{\omega}_{\mu})e^{-S_{0}(A^{\omega}_{\mu})+\int\bar{c}
\sqrt{-\partial_{\mu}\partial_{\mu}}cdx}\nonumber\\
&&=\int{\cal D}A^{\omega}_{\mu}{\cal D}B{\cal D}\bar{c}{\cal D}c 
e^{-S_{0}(A^{\omega}_{\mu})+\int[-iB\frac{1}{\sqrt{-\partial_{\mu}
\partial_{\mu}}}\partial_{\mu}A^{\omega}_{\mu}+\bar{c}
\sqrt{-\partial_{\mu}\partial_{\mu}}c]dx}\nonumber\\
&&=\int{\cal D}A^{\omega}_{\mu}{\cal D}B{\cal D}\bar{c}{\cal D}c 
e^{-S_{0}(A^{\omega}_{\mu})+\int[-iB\partial_{\mu}A^{\omega}_{\mu}+\bar{c
}(-\partial_{\mu}\partial_{\mu})c]dx}
\end{eqnarray}
after the field re-definition of auxiliary variables $B$ and 
$\bar{c}$ in (3.13).
This procedure is the one we used for the non-Abelian
case. 

\section{Conclusion}
We have shown explicitly the equivalence of the modified path
integral formula\cite{zwanziger}\cite{jona-lasinio} to the 
conventional Faddeev-Popov formula\cite{faddeev} without taking 
any limit of the gauge fixing parameter, if the Gribov 
complications are absent. In the presence of Gribov
complications, these two formulas are equivalent but not identical
to each other, and both give rise to non-local theory in general 
and thus the unitarity is not obvious. 

From a view point of non-perturbative definition of gauge
theory, a BRST invariant formulation of lattice gauge theory
is important\cite{golterman}. The Neuberger's stricture in the BRST 
invariant 
lattice formulation\cite{neuberger} corresponds to the appearance of an 
{\em even}
number of copies with alternating signature of 
$det\{\frac{\delta}{\delta\omega}
[D_{\mu}(\frac{\delta f(A^{\omega}_{\nu})}{\delta A^{\omega}_{\mu}})]\}$
due to the periodicity of lattice gauge variables, and thus
(when we use the continuum notation in (2.28))
\begin{equation}
\int{\cal D}A_{\mu}^{\omega}\delta (D_{\mu}(\frac{\delta 
f(A^{\omega}_{\nu})}{\delta A_{\mu}^{\omega}}))det\{\frac{\delta}{\delta\omega}
[D_{\mu}(\frac{\delta f(A^{\omega}_{\nu})}{\delta A^{\omega}_{\mu}})]\}
e^{-S_{0}(A^{\omega}_{\mu})}=0.
\end{equation}
The recent analysis by Testa\cite{testa} suggests that the BRST 
invariant 
ansatz\cite{fujikawa} can  be implemented for Abelian theory and also for 
the Abelian projection of non-Abelian theory by a suitable 
generalization of the $\delta$-function on the lattice. 
It remains an interesting problem to extend the analysis of 
Testa to fully non-Abelian lattice theory which may eventually
overcome the Neuberger's stricture. 

As for the practical implications of the present continuum
 analysis on the lattice simulation, it may be important to 
remember that the gauge 
fixing by adding an (effective) mass term to the lattice action
\cite{golterman} may not go beyond the conventional 
Faddeev-Popov procedure if one accounts the variation of the 
gauge variable along the entire gauge orbit. This property will 
be important  when one analyzes the issues related to  unitarity.
   It should also be emphasized that both of the original 
Faddeev-Popov
formula and the modified one give rise to a positive definite 
integrand in the path integral even on the 
lattice\cite{golterman}.
\\

The present note was motivated by the discussions at the RIKEN-BNL
Workshop, May 25-29, '99 and at the NATO Advanced Research
Workshop at Dubna, October 5-9, '99. One of the authors (KF) 
thanks all the 
participants of those workshops for stimulating discussions.

\end{document}